\documentstyle[12pt,eqsecnum,multicol,epsfig,aps,prl,array]{revtex}
\newcommand{\bleq}{\ifpreprintsty
   \else
   \end{multicols}\widetext \vspace*{-3.5ex}{\tiny
   
\noindent\begin{tabular}[t]{c|}
   \parbox{0.493\hsize}{~} \\ \hline \end{tabular}}
      \fi}
\newcommand{\eleq}{\ifpreprintsty
   \else
   {\tiny\hspace*{\fill}\begin{tabular}[t]{|c}\hline
    \parbox{0.49\hsize}{~} \\
    \end{tabular}}\vspace*{-2.5ex}\begin{multicols}{2}
    \narrowtext
    \fi}
\newcommand{\bcols}{\ifpreprintsty\else\begin{multicols}{2} 
\narrowtext\fi}
\newcommand{\ecols}{\ifpreprintsty\else\end{multicols}\fi}

\draft
\widetext
\begin{document}
\title{Percolative conductivity in alkaline earth silicate melts and glasses} 
\author{M. Malki$^{\dag}$, M. Micoulaut$^{\ddag}$, F. Chaimbault$^{\dag}$, 
Y. Vaills$^{\dag}$, P. Simon$^{\dag}$}

\address{
$^{\dag}$ CRMHT-CNRS 1D, avenue de la Recherche Scientifique\\
45071 Orl{\'e}ans Cedex 02, France\\
Universit{\'e} d{\'{}}Orl{\'e}ans, 45067 Orl{\'e}ans Cedex 02, France
\ \\
$^{\ddag}$ Laboratoire de Physique Th{\'e}orique des Liquides
Universit{\'e} Pierre et Marie Curie\\  Boite 121
4, Place Jussieu, 75252 Paris Cedex 05, France\\}

\date{\today}
\maketitle
\begin{abstract}
\par
Ion conducting $(CaO)_x(SiO_2)_{1-x}$ glasses and melts show a
threshold behaviour in dc conductivity near $x=x_t=0.50$, with
conductivities increasing linearly at $x>x_t$. We show that the
behaviour can be traced to a rigid ($x<0.50$) to floppy ($x>0.50$)
elastic phase transition near $x=x_t$. In the floppy phase,
conductivity enhancement is traced to increased mobility or diffusion
of $Ca^{2+}$ carriers as the modified network elastically softens.
\par
{Pacs:} 61.43Fs-62.20.-x
\end{abstract}
\newpage
Electrical and thermal properties are directly related to the
structure of silicate melts and investigations 
into their structure have received therefore considerable
attention in earth science. It is indeed of fundamental importance 
to understand how
magmas which contain alkali and alkaline earth silicates behave under 
different physical circumstances. Alkali silicate glasses have been 
extensively studied within this context and numerous articles have
focused on the variation of structure\cite{r1}, conductivity\cite{r2}, 
glass transition\cite{r3}, etc. with respect to 
the alkali concentration.
On the other hand, a small number of studies on alkaline earth silicate 
glasses have been reported to date. Some of them, however, have focused
either on the structure of glasses containing barium and calcium 
silicates\cite{r5}, or on the thermodynamics and 
miscibility of the corresponding melts\cite{r6} and density or
specific volume\cite{r6b}. It is worth mentioning
that most of these studies have been restricted around the $50\%$
concentration of alkaline earth oxide where glass-forming tendency (GFT)
is optimized\cite{r7}. Finally, even though most of the fast ionic conductors 
(FIC)\cite{FIC1,FIC2} 
having potential applications in electrochemical devices such as solid
state batteries are alkali (or silver) silicates or thiosilicates, the
conductivity of alkaline earth silicates is non-negligible [about
$2.10^{-7}~\Omega^{-1}.cm^{-1}$ at $550^oC$] and the role of the migrating
calcium cations
remains to be completely understood. Our work attempts to address for
the first time this basic issue.
\par
In this Letter, we report on $(CaO)_x(SiO_2)_{1-x}$ glasses and the
common physical origin for the
behavior in both the GFT and the conductivity data that follow
directly from the Phillips-Thorpe constraint theory\cite{r8}. 
Calcium silicate
melts and glasses near $x=0.5$ are optimally constrained and display
percolative conduction in both melts and glasses. This is a new
feature that has never been observed in the corresponding alkali
silicates, which links ionic conductivity to glass elasticity and
structure. Furthermore, we show that the sudden
increase of the conductivity at high calcium concentration has to be a
consequence of the calcium cation mobility percolation arising from the
floppy to rigid transition\cite{r9}. The major consequence is that
here the carrier concentration does not dominate the conductivity, as
usually reported.
\par
The samples were prepared by mixing pre-dried $SiO_2$ (99.99 $\%$) and
$CaCO_3$ (99.95 $\%$) powders in the correct proportions. For each
sample, the mixture was melted in a platinum crucible at $1650^{o}$C for 
two hours, and quenched by plunging the bottom of the crucible in cold
water. The glass 
transition temperatures were determined with a differential scanning 
calorimeter Setaram DSC-1600 at a heating rate $10^{o}C/min$. These
values are slightly larger than those reported by Shelby\cite{Shelby}
using the dilatometric technique but they exhibit the same 
global trend (fig. 1), i.e. a plateau up to a Calcium concentration
in the range [0.45-0.48] and a more or less linear increase for higher
concentrations. In the solid state, the complex electrical
conductivity was measured with
an impedance spectrometer 4194A in the temperature range
$500^{o}C-950^{o}C$ and in the frequency range 100 Hz-10 MHz. The
samples were $0.5~cm^2$ glass pellets with a thickness about
$1~mm$. Platinum was evaporated as electrodes on both faces of the
pellet. The temperature was measured with a Pt/Pt-$10\%$Rh thermocouple
located at about $1~mm$ from the sample\cite{Malki}.  
We restricted our study in the molten state
to only three compositions.
\par
The dc conductivity $\sigma$ displays almost the same trend as $T_g(x)$, with a very
small value (about $10^{-7}~\Omega^{-1}.cm^{-1}$) for $x<0.47$ and a sudden increase
for larger compositions. The almost constant
value of $\sigma$ in the low calcium region and the sudden increase for
$x>0.47$ has been also obtained in simulations and experiments for the
molten state (fig. 2, insert\cite{computer}), suggesting that molten
and glassy state behave similarly. Furthermore, we have observed the same 
kind of trend in our molten samples (see fig. 3, below) which are
almost at the same temperatures than those of Ref. \cite{computer}.
Figure 3 shows
the Arrhenius plots [$\sigma T=\sigma_0\exp[-E_A/k_BT]$] of the conductivity $\sigma$ for
the glassy and the molten state. 
From
the latter, it is also obvious that the conductivity data 
of three compositions in the $x\leq 0.50$ range map onto each other and have the
same global trend in the glass.
\par
These results can be quantitatively understood using the tools of
Lagrangian bonding constraints as introduced by Phillips\cite{r8}.
A network constrained by bond-stretching and bond-bending forces sits
indeed at a mechanically critical point when the number of constraints per
atom $n_c$ equals the network dimensionality. At this point, glass optimum
is attained and GFT is enhanced\cite{r13b}. These ideas can be cast in terms
of percolation theory by evaluating the number of zero frequency modes (floppy
modes $F/N$) from a dynamical matrix\cite{r9}, which vanish at the
floppy to rigid transition. The very accurate agreement of
experiments in chalcogenides with the predictions of the theory is quite
remarquable\cite{r15,r16} but further investigations devoted to
alkali oxide glasses have been performed only recently\cite{Science,RK}.
\par
We consider the $CaO-SiO_2$ system as a network of $N$ atoms
composed of $n_r$ atoms that are $r-$ fold coordinated. Enumeration of
mechanical 
constraints gives for bond-stretching forces $r/2$ and for 
bond-bending forces $(2r-3)$ constraints. The average number of floppy
modes per
atom $F/N$ in a three-dimensional network is given by\cite{r9}:
\begin{eqnarray}
\label{1}
F/N&=&n_d-n_c=3-{\frac {1}{N}}\sum_{r\geq 2}n_r({\frac {5r}{2}}-3)
\end{eqnarray}
which applied to the present $(CaO)_x(SiO_2)_{1-x}$ case, yields:
\begin{eqnarray}
\label{2}
F/N&=&3-{\frac {11-7x}{3-x}}
\end{eqnarray}
provided that $Si$ is four-folded and calcium and oxygen are two-fold 
coordinated. The coordination number used in equ. (\ref{2}) for $Ca$ desserves
some comments since computer simulations\cite{computer} and EXAFS
studies\cite{EXAFS} have suggested that the number of nearest neighbors
of the calcium atom may be about 6. However, Debye-Waller factor in
these studies on combined
systems using either $BaO$ or $CaO$ as network modifier are slightly
different, suggesting 
that the oxygen neighbors of $Ca$ were not all equivalent\cite{EXAFS} thus
promoting the chemical hypothesis for the coordination number of Ca ($r=2$).
On the other hand, in constraint theory one deals only with the
bonds that provide a strong mechanical constraint, which means that the
covalent bonds are the most qualified. To be more specific, one can use
Pauling's definition to determine the fractional ionic character $f$ in
the Ca-O bond\cite{Pauling}, which yields here the value
$f=0.774$ and a corresponding covalency factor of 0.226. With a
coordination of 6 for $Ca$, the corresponding covalent coordination is
$1.36$, somewhat lower than $2$. Furthermore, mechanical constraints
provided by alkali atoms
connected to more than one oxygen have been suggested to be only resonating
constraints\cite{RK}, which do not participate in the global counting. 
Finally, alkali silicate and tellurate glasses are illustrative
precedents\cite{Science}. In these systems, the floppy to rigid transition 
occurs at the expected critical concentration\cite{Yann} only if the
involved atoms (Na) have a covalent coordination number of $1$ (and
the ``Pauling{\"{}} analysis\cite{indi} yields the covalent coordination 
$0.832-1.04$), even though the number of nearest
neighbors\cite{Greaves} of the sodium atom is believed to be 5. 
\par
Our interpretation of the dc conductivity is as follows. At $x=x_c=0.5$, 
the number of floppy modes $F/N$ vanishes and the network undergoes a
rigid to floppy transition, following equ. (\ref{2}). For $x<0.5$ 
the system is stressed rigid, i.e. there are more constraints than
degrees of freedom. As a consequence,
the mobility of the calcium cation is very weak because the cation has to
overcome a strong mechanical deformation energy to move from one
anionic site to another. In an ideal floppy glass at
$x>0.5$ where only bond-bending and bond-stretching forces are 
considered\cite{r8} this deformation energy is zero and related 
elastic constants ($c_{11}$, $c_{44}$)
vanish\cite{Tersoff}. Therefore, percolation of floppiness equals 
percolation of $Ca$ mobility. As a
result, one has a substantial
increase of the mobility hence of the conductivity. In the
strong
electrolyte Anderson-Stuart model\cite{Anderson}, the activation energy for
conductivity
$E_A=E_C+E_m$ depends on the Coulombic term $E_C$ (which controls the
carrier rate) and a strain term
$E_m$ related to the mobility  that can
be thought as the energy required to enlarge the radius of an anionic
site of length $l(x)$ by an amount $\delta r$, the latter quantity being
roughly equal to the cation radius:
\begin{eqnarray}
\label{strain}
E_m(x)&=&{\frac {1}{2}}\pi c_{44}(x)l(x)(\delta r)^2
\end{eqnarray}
The energy term $E_m(x)$ depends on the shear modulus $c_{44}(x)$ which in
turn depends on the Ca concentration\cite{Barrio}. In a floppy model
network, the shear modulus $c_{44}(x)$ is zero\cite{Tersoff} leading to a zero
value for the strain energy $E_m(x)$ and a maximum in mobility.
\par 
Additional support for our interpretation derives from two other
observations: (1) In
the molten $(CaO)_x-(SiO_2)_{1-x}$, we still observe an increase 
in conductivity with $x$ but with values which are substantially higher
compared to the glass (fig. 3) and
which can be fitted with a Vogel-Fulcher-Tamman (VFT) law:
\begin{eqnarray}
\label{l4}
\sigma T=Aexp[-\Delta/(T-T_0)]
\end{eqnarray}
arising from the variation in temperature of
the calcium diffusion constant (via the Nernst-Einstein 
equation)\cite{Anderson}. From the
fits (solid lines in insert of fig. 3) it appears that the 
pseudo-activation energy $\Delta$ is constant for
the two values in the floppy region ($\Delta=2733.0$ and $2740.7 K$ for
respectively $x=0.44$ and $x=0.5$) while it increases for the
composition 
$x=0.53$ ($\Delta=3008~K$). Also, the VFT temperature $T_0$ at which the 
diffusion constant (and
the underlying relaxation time towards thermal equilibrium) diverges
is dynamically inaccessible during glass transition, but it is
accepted that its value is close to the ideal Kauzmann\cite{Kauzm}
glass transition temperature $T_K$. An ideal glass would therefore
have its $T_g$ close to $T_K$ and $T_0$, a situation which is met in the present
system for the Phillips' optimal glass composition $x=0.5$
($T_g/T_0=1.0002$ compared to $T_g/T_0=1.0218$ for $x=0.44$ and
$T_g/T_0=1.0621$ for $x=0.53$).
\par
In fig. 4  are represented the Arrhenius activation energy $E_A$ for 
conductivity and the preexponential factor $\sigma_0$ as a
function of the mean coordination number of the network\cite{r9}
which is $\bar r=(8-4x)/(3-x)$. The former displays a minimum at
$\bar r=2.4$ which corresponds to $x=0.5$. 
\par
This in turn can be compared to the 
non-reversing heat flow  extracted from complex calorimetric
measurements at the glass transition temperature for various
chalcogenide systems\cite{Bool}.
Here, this relaxing part of the total heat flow exhibits
always a minimum at the floppy to rigid transition (close to the
mean-field\cite{r8} value $\bar r=2.4$). 
The same global trend 
is observed in numerous chalcogenides\cite{Bool}, 
and our calcium silicate glasses exhibit also a minimum in $E_A$ at
the mean coordination number $\bar r=2.4$. 
From the chalcogenide example, we suggest that when
$x=0.5$ structural relaxation of the glass network proceeds with
minimal enthalpic changes because the minuscule network stress is
uniformly spread when the network is mechanically critical 
(i.e. F/N=0). Furthermore, we note that the variation of $\sigma$ is here
not driven by the compositional trends of $\sigma_0$. It appears to be
rather a balance between the variation of $E_A$ and $\sigma_0$.
(2) Rigidity percolation thresholds in Micro-Raman
measurements performed in backscattering geometry have been reported
in Ref. \cite{r15}. Therefore, 
we have studied in the present work as a function of the Ca
concentration the frequency
$\nu$ and the linewidth $\Gamma$ of the stressed rigid $SiO_{4/2}$ ($Q^4$
unit, where the superscript denotes the number of network bridging
oxygens)). The corresponding line ($A_1$ stretching mode) 
exhibits a change of regime (fig. 5) for the line frequency
at the concentration $x=0.50$, consistently with chalcogen analogs\cite{r15}. 
We concentrate here only on this (stressed rigid) line and will report
separately the complete Raman analysis elsewhere\cite{Raman}. On the
other hand, the 
evolution of the linewidth $\Gamma$ with $x$ permits to follow the local 
environnement of the $Q^4$ unit. For $x<0.50$, $\Gamma$ remains constant,
 related to the absence of change in the coupling of this unit with
the rest of the network. It is the coupling which makes possible the
presence of rigid regions (through 
isostatic $Q^4-Q^3$ and stressed $Q^4-Q^4$ bondings)
although the number of floppy $Q^2$ and $Q^1$ units is steadily increasing. 
Above the critical concentration $x=0.5$, the sharp drop of $\Gamma$ clearly shows 
decoupling of the $Q^4$ unit with respect to the network (fig. 5),
signifying decoupling of stressed rigid regions thus percolation of
floppiness.
\par
In FICs, it is often believed that it is the carrier concentration
that dominates the conductivity. In the strong electrolyte 
model\cite{Anderson}, the cation electostatic Coulombic energy
barrier has to be overcome to ensure conduction. On the other hand, in
the weak electolyte model a dissociation energy is needed to
create the mobile carrier\cite{weak}. These two pictures remain of course valid
as long as the sizes of the cations is weak compared to the
interstices of the glass network.
\par
Our conclusion  brings us back to the analogy with alkali
oxide FIC's and the popular conductivity channel picture\cite{Greaves}. In 
these glasses, the rigid to floppy transition occurs
at the concentration $x_c=0.20$ which is very close to the reported
threshold concentration separating intrachannel cation hopping
(involving a weak mechanical deformation of the network, since the
motion occurs only in macroscopic holes of the network) from
network hopping (strong mechanical deformation only possible in a floppy
network). However, in alkali silicates, no typical behavior emerges in
compositional trends of the conductivity because of the 
growing contribution of the free
carrier rate\cite{Greaves}. This has to be put in contrast with our
present study on calcium silicates
where the conductivity is almost constant up to a critical value $x_c\simeq
0.48$ beyond which $\sigma$ steadily increases. Therefore, the free carrier
concentration cannot be considered as an increasing function of 
alkaline earth
composition. We believe that
conductivity in calcium silicates is driven by the carrier mobility in
the network and percolates at the rigid to floppy transition.
On this system, further investigations and measurements of physical quantities
displaying usually a threshold behavior\cite{rigid} at the rigidity transition
will be achieved in close future.
\par LPTL is Unit{\'e} Mixte de Recherche n. 7600. It is a pleasure to
acknowledge the help of Boris Robert during the course of this work.

\newpage
\vspace{0.8cm}
\begin{figure}
\begin{center}
\epsfig{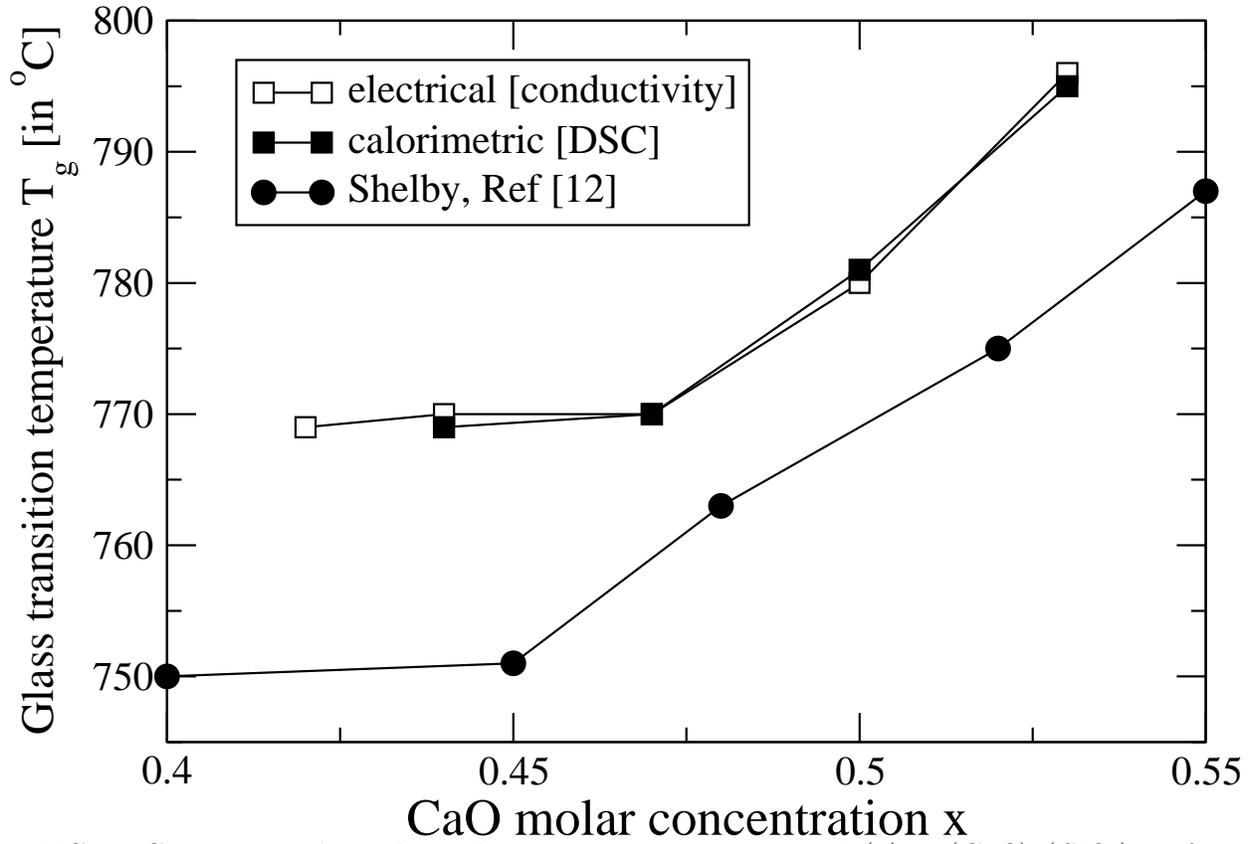}
\caption{Compositional trends in glass transition temperature $T_g(x)$ 
in $(CaO)_x(SiO_2)_{1-x}$ from DSC and electrical measurements,
compared to previous studies[12]}.
\end{center}
\end{figure}
\newpage
\vspace{0.5cm}
\begin{figure}
\begin{center}
\epsfig{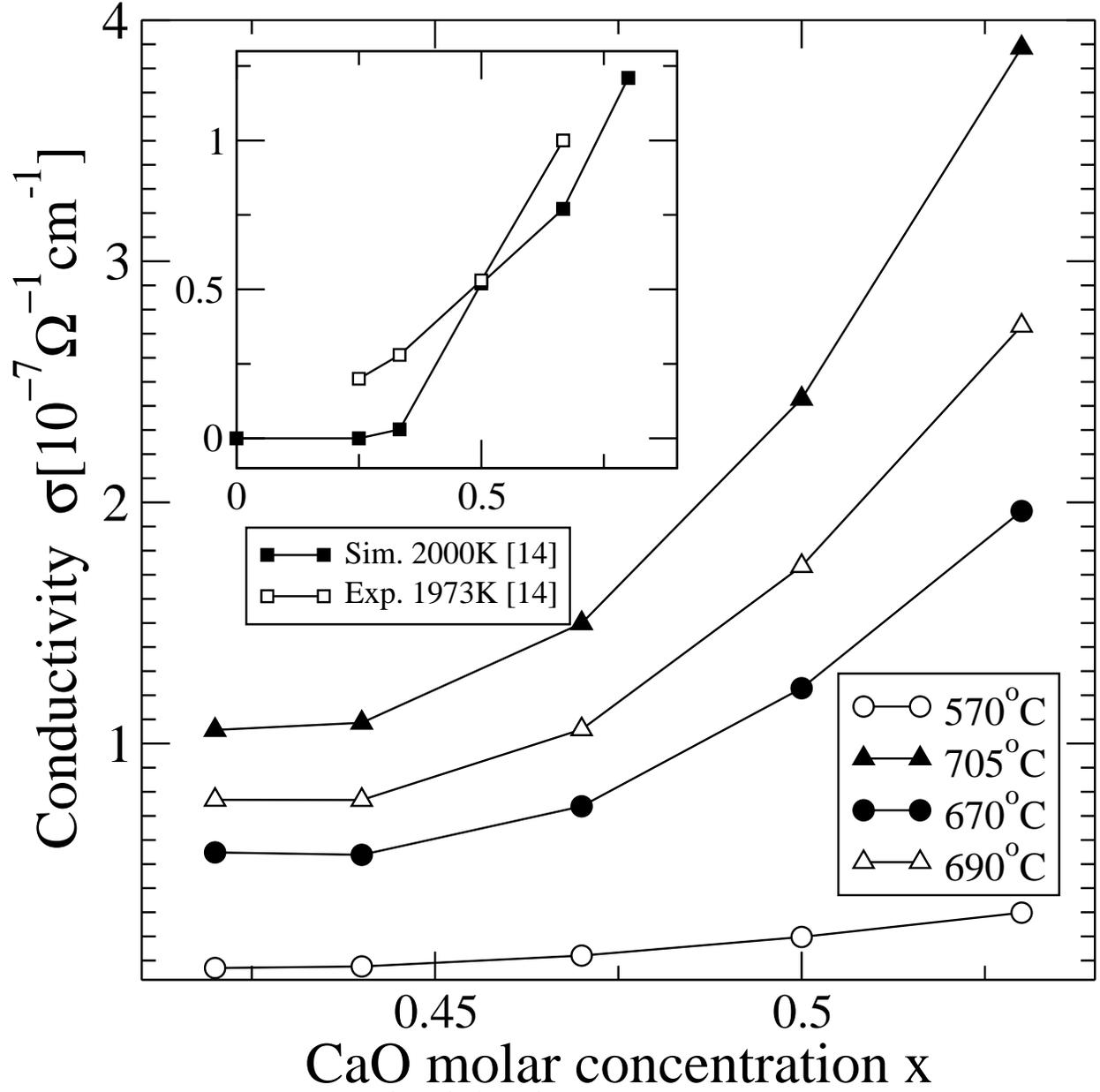}
\caption{dc conductivity $\sigma$ in $(CaO)_x(SiO_2)_{1-x}$
glasses with $Ca$ composition $x$ for different temperatures. The
insert shows the conductivity in the melt from simulation and previous
experiments[14] at around $2000~K$.}
\end{center}
\end{figure}
\newpage
\vspace{0.5cm}
\begin{figure}
\begin{center}
\epsfig{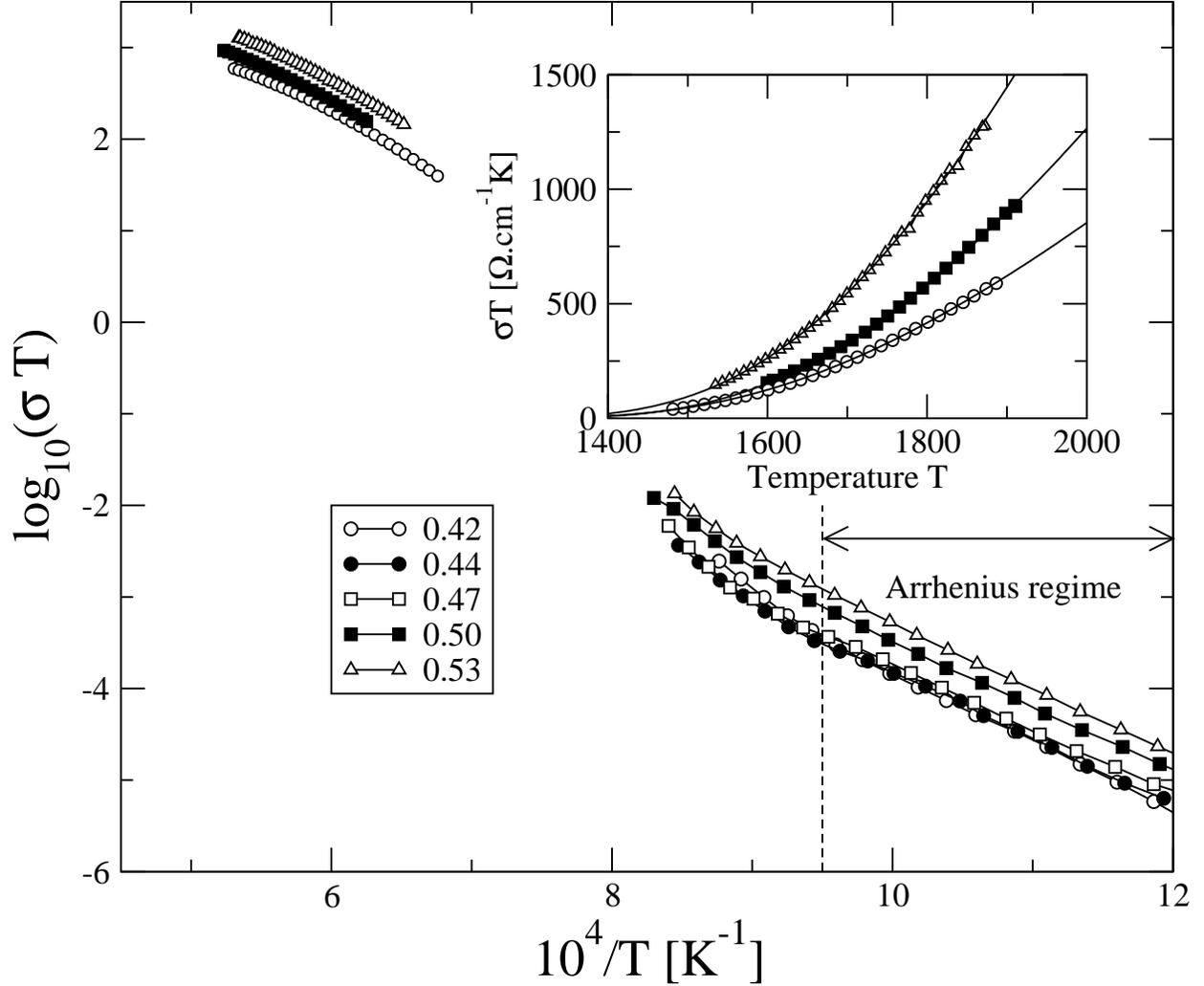}
\caption{Arrhenius plots of the ionic conductivity for
$(CaO)_x(SiO_2)_{1-x}$ glasses in the solid and the molten state. The
fit for the estimation of $E_A$ and $\sigma_0$ have been performed at low
temperature in the Arrhenius regime. Note
the small 
deviation occuring close to the glass transition temperature. The
insert shows $\sigma T$ in the liquid as a function of temperature with
corresponding Vogel-Fulcher-Tamman (VFT) fits.}
\end{center}
\end{figure}
\newpage
\begin{figure}
\begin{center}
\epsfig{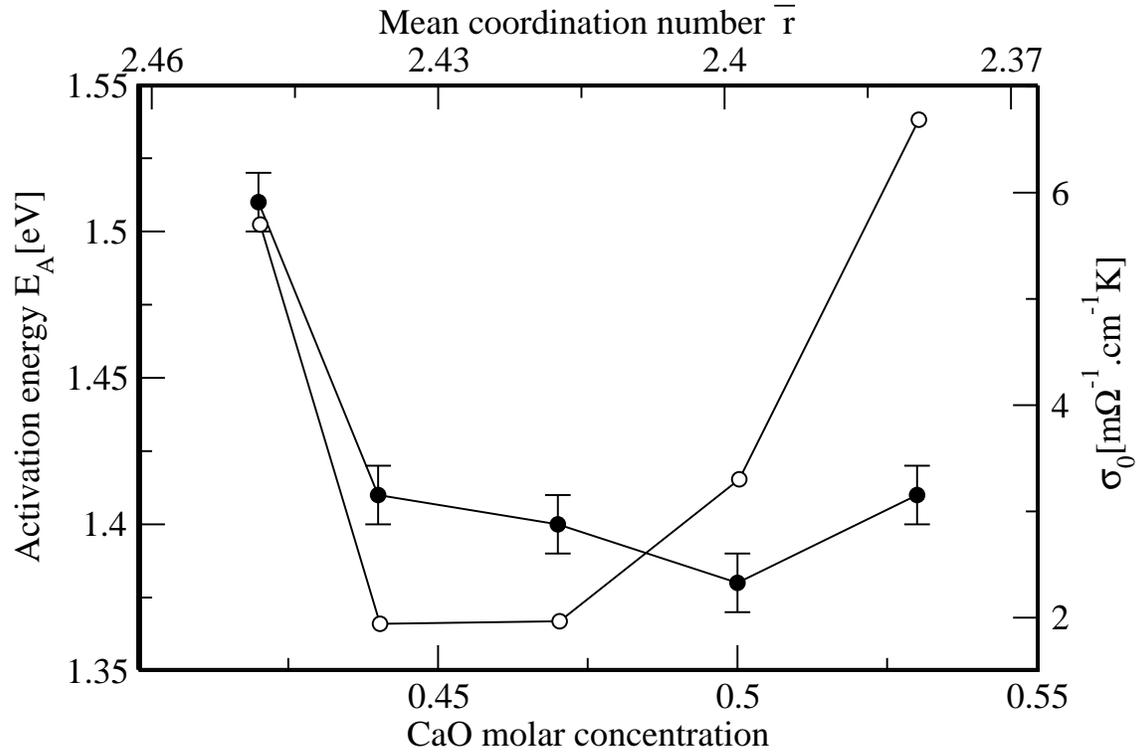}
\end{center}
\caption{Arrhenius parameters (activation energy $E_A$ (solid circles)
and preexponential factor $\sigma_0$ (open circles)) as a
function of the mean coordination number $\bar r$ in the glass phase.}
\end{figure}
\newpage
\vspace{0.8cm}
\begin{figure}
\begin{center}
\epsfig{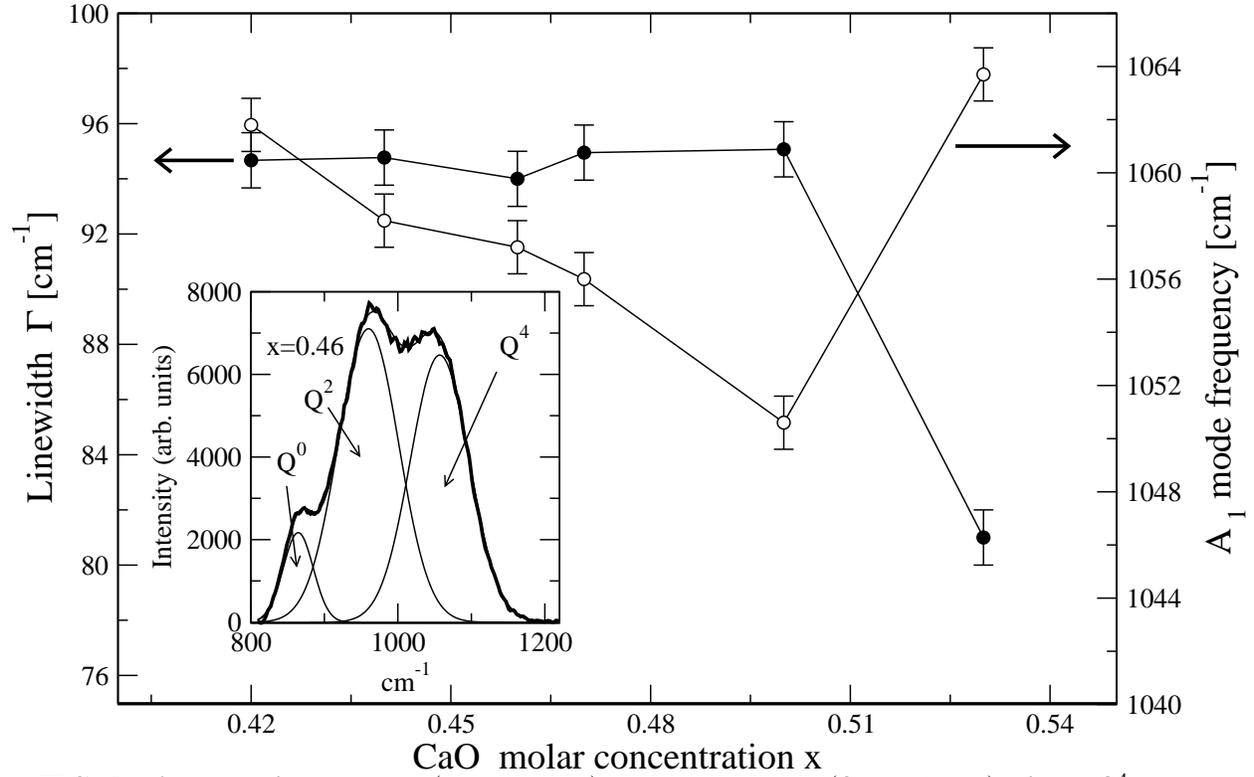}
\caption{$A_1$ mode frequency $\nu$ (open circles) and linewidth $\Gamma$
(filled circles) of the $Q^4$ line as a
function of Ca concentration. The insert shows part of the Raman
spectra with the line of interest for 
$x=0.46$ and the related Gaussian fits at $850~cm^{-1}$ [$Q^0$ unit] and
$950~cm^{-1}$ [$Q^2$ unit] from [32].}
\end{center}
\end{figure}

\end{document}